\documentclass[aps,
groupedaddress,showpacs,
nofootinbib,twocolumn]{revtex4}
\usepackage{amssymb,amsmath}
\usepackage{graphicx}
\usepackage[english]{babel}
\usepackage[usenames,dvipsnames]{color}

\newcommand{\ket}[1]{\left | #1 \right \rangle}
\newcommand{\bra}[1]{\left \langle #1 \right |}
\newcommand{\braket}[2]{\langle#1|#2\rangle}
\newcommand{\proj}[1]{\ket{#1}\bra{#1}}
\newcommand{\avg}[1]{\langle#1\rangle}
\newcommand{\bigavg}[1]{\big\langle#1\big\rangle}

\definecolor{darkred}{rgb}{.8,0,0}

\definecolor{darkblue}{rgb}{0,0,.7}

\newcommand{\oline}[1]{\overline{#1}}

\begin{document}

\title{Local time of L\'{e}vy random walks: a path integral approach}

\author{V\'{a}clav Zatloukal}
\email{zatlovac@gmail.com}

\affiliation{Department of Physics, Faculty of Nuclear Sciences and Physical Engineering,\\
Czech Technical University in Prague, B\v{r}ehov\'{a} 7, 115 19 Praha 1, Czech Republic}

\begin{abstract}
Local time of a stochastic process quantifies the amount of time that sample trajectories $x(\tau)$ spend in the vicinity of an arbitrary point $x$. For a generic Hamiltonian, we employ the phase-space path-integral representation of random walk transition probabilities in order to quantify the properties of the local time. For time-independent systems, the resolvent of the Hamiltonian operator proves to be a central tool for this purpose.
In particular, we focus on local times of L\'{e}vy random walks (L\'{e}vy flights), which correspond to fractional diffusion equations.
\end{abstract}

\pacs{05.40.-a, 02.50.Ga, 05.40.Fb, 31.15.xk}


\maketitle

\section{Introduction}

We consider a diffusion equation (also referred to as the heat or the Fokker-Planck equation) of the form
\begin{equation} \label{HeatEq}
\left[ \partial_t + H(-i \partial_x,x,t) \right] P(x,t)
= 0 ,
\end{equation}
where $H(p,x,t)$ is a generic Hamiltonian, and $x \in \mathbb{R}$.
Its solution for the initial condition $P(x,t_a) = \delta(x-x_a)$ can be represented via the phase-space path integral \cite{KleinertPI}:
\begin{equation} \label{CondProbPI}
(x_b t_b | x_a t_a)
\equiv P(x_b,t_b)
= \int\displaylimits_{x(t_a)=x_a}^{x(t_b)=x_b} \hspace{-3mm} \mathcal{D}x \int \frac{\mathcal{D}p}{2\pi}\, e^{\mathcal{A}[p,x]} ,
\end{equation}
where 
\begin{equation} \label{Action}
\mathcal{A}[p,x]
= \int_{t_a}^{t_b} \!\!d\tau \big[ i p(\tau) \dot{x}(\tau) - H(p(\tau),x(\tau),\tau) \big] 
\end{equation}
is the action corresponding to the Hamiltonian $H$. When tractable, the functional integration over the momentum paths $p(\tau)$ yields a configuration-space path-integral representation, commonly referred to as the Feynman-Kac formula \cite{Kac1949} (see Ref.~\cite{Eule2014} for its generalization).

The path integral point of view suggests to interpret the quantity~\eqref{CondProbPI} as (possibly unnormalized) transition probability, that is, the conditional probability that a particle localized at an initial time $t_a$ at point $x_a$ will be found at a later time $t_b$ at point $x_b$.

Let us note that for imaginary times, Eq.~\eqref{HeatEq} becomes the Schr\"{o}dinger equation. Eq.~\eqref{CondProbPI} then represents the transition amplitudes, i.e., the matrix elements of the evolution operator, thus constituting the path-integral formulation of quantum mechanics \cite{Dirac1933,Feynman1948}. In this article, however, we will only consider diffusion processes described by Eq.~\eqref{HeatEq} in real time domain. The respective path (or stochastic) integral was first studied by Wiener \cite{Wiener1923} in order to mathematically describe the Brownian motion. In fact, the path integrals of Eq.~\eqref{CondProbPI} find applications in quantum physics too, namely, they represent the density matrix of a system in thermal equilibrium \cite{KleinertPI}.

The position distribution of a Brownian particle, i.e., the Gaussian (or normal) distribution, obeys the standard diffusion equation, which is obtained by taking the Hamiltonian $H \propto p^2$.
The latter is a member of a broader class of Hamiltonians,
\begin{equation} \label{LevyHam}
H_\lambda(p) = D_\lambda (p^2)^{\lambda/2} ,
\end{equation}
labeled by a real parameter $\lambda$, which will, in this article, take values in the interval $[1,2]$. The diffusion equation~\eqref{HeatEq} then features a fractional Laplace differential operator \cite{Laskin2002,Podlubny}, and is therefore sometimes called fractional diffusion (or Fokker-Planck) equation \cite{KleinertZatl2013}. Its fundamental solution $P_\lambda(x,t)$ has the form of a L\'{e}vy stable distribution \cite{Levy1925,Sato}. These probability distributions are characterized by a typical power-like heavy-tail behaviour for $\lambda < 2$, while for $\lambda=2$ they reduce to the Gaussian distribution. 
Their prominent feature is the fact that they occur as the limiting distributions in the generalized central limit theorem \cite{GnedenkoKolmog}.

L\'{e}vy random walks (or stable stochastic processes \cite{Ito}) correspond to a path integral with L\'{e}vy Hamiltonian $H_\lambda$. Their sample paths exhibit long jumps, the so-called L\'{e}vy flights, which become more pronounced for decreasing values of the parameter $\lambda$ (see Fig.~\ref{fig:SamplePaths} below). This is due to accumulation of the probability of large steps in the heavy tails of the L\'{e}vy distributions.

L\'{e}vy statistics is pertinent in cases when the standard Gaussian statistics is inadequate, typically when the random variable has infinite variance.
It finds numerous applications in physics (e.g., in anomalous diffusion \cite{Bouchaud1990},
laser cooling \cite{Bardou2002}, etc.), but also in finance \cite{KleinertPI,MantegnaStanley} and other fields.

The notion of \emph{local time} of a stochastic process (cf. also the \emph{occupation} or \emph{sojourn time}) was introduced by L\'{e}vy \cite{Levy1939} as a measure of time that a stochastic trajectory $x(\tau)$ spends in the vicinity of an arbitrary point $x$ (see monograph~\cite{MarcusRosen} for an overview). Formally, one defines the density
\begin{equation} \label{LTdef}
L(x;t_a,t_b,x(\tau)) = \int_{t_a}^{t_b} \!\!d\tau \,\delta(x-x(\tau)) ,
\end{equation}
which is a non-negative random variable indexed by $x$. The functions $L(x)$ are normalized to $t_b-t_a$, moreover, it has been shown that they are continuous for a broad class of processes including L\'{e}vy random walks with $\lambda>1$ \cite{Boylan1964}. 

To any path $x(\tau)$ there corresponds a local-time profile $L(x)$, which can in turn be regarded as a sample trajectory of a new stochastic process. Notably, for the case of Brownian motion, the latter has been identified as a squared Bessel process \cite{Ray1963,Knight1963,Borodin1989}. The ensuing local-time path-integral representation \cite{JizbaZatl2015} provides an alternative to the standard Feynman path integral, especially illuminating for long evolution times, which correspond to the low-temperature regime in quantum statistical physics \cite{Paulin2007}. Among other applications of local times in physics, let us mention, for example, quantum scattering or tunneling processes \cite{Sokolovski1987}.
\begin{figure*}
\includegraphics[width=17cm]{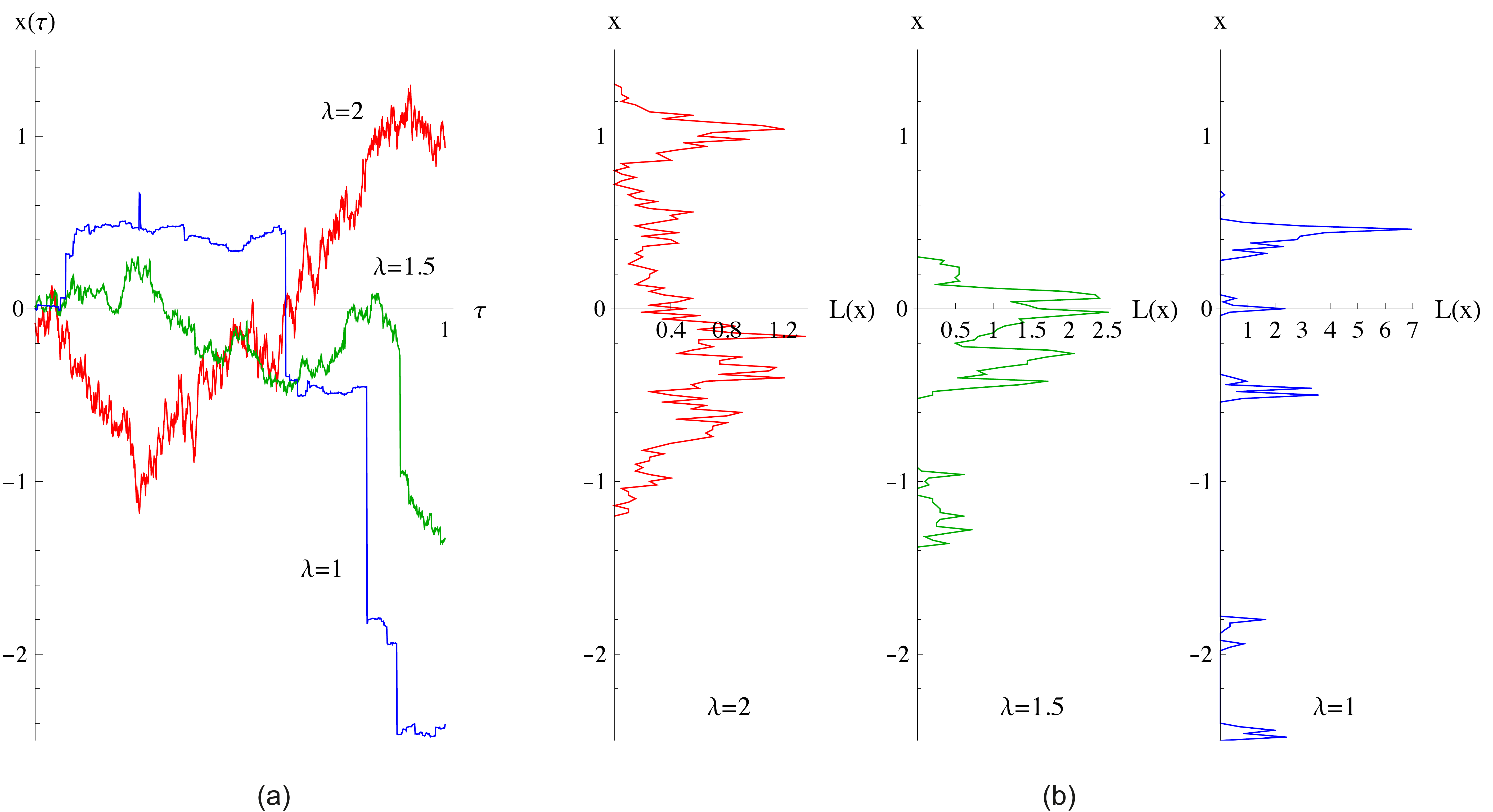} 
\caption{(a)~Sample paths $x(\tau)$ of a L\'{e}vy random walk for $\lambda=2$ (Brownian motion), $\lambda=1.5$, and $\lambda=1$ (Cauchy walk). (b)~Local-time profiles $L(x)$ corresponding to the sample paths $x(\tau)$.}
\label{fig:SamplePaths}
\end{figure*}

In this article, methods of path integration are employed to probe some properties of local times of random walks described by generic Hamiltonians. In Sec.~\ref{sec:CorrFunFunc} we derive basic formulas for the correlation functions, or even generic functionals of the local time. These are exploited further in Sec.~\ref{sec:TimeIndep} under the assumption that the Hamiltonian does not explicitly depend on time, and turned into a practical computational tool by performing the Laplace transform in the time variable, and using some properties of resolvent operators. In Sec.~\ref{sec:Levy} we consider specifically the L\'{e}vy Hamiltonian~\eqref{LevyHam}, and provide explicit results for one-point distributions and averages of the local time. Multipoint distributions and higher moments and are discussed in Appendices~\ref{sec:MultiDist} and \ref{sec:HigherMom}, respectively.

\section{Correlation functions and functionals of the local time}
\label{sec:CorrFunFunc}

Correlation (or $n$-point) functions of the local time are the functional averages
\begin{equation} \label{CorrFuns}
\avg{L(x_1) \ldots L(x_n)}
=\! \hspace{-2mm} \int\displaylimits_{x(t_a)=x_a}^{x(t_b)=x_b} \hspace{-3mm} \mathcal{D}x \int \frac{\mathcal{D}p}{2\pi}\, e^{\mathcal{A}[p,x]} L(x_1) \ldots L(x_n) .
\end{equation}
They define moments of the local time
\begin{equation} \label{Moments}
\mu(x_1,\ldots,x_n) 
= \frac{\avg{L(x_1) \ldots L(x_n)}}{\avg{1}} ,
\end{equation}
where the normalization factor ${\avg{1}}$ is equal to $(x_b t_b | x_a t_a)$. [For the sake of brevity, the dependence of $\mu$ on $t_a$, $x_a$, $t_b$ and $x_b$ is not displayed.]

The path integral of Eq.~\eqref{CorrFuns} is summing over paths with fixed initial and end point. Later on, we shall also consider the set of trajectories starting at $x_a$ and ending anywhere, in which case we denote
\begin{equation} \label{EndpointInt}
\avg{\,\ldots\,}^{*} 
\equiv \int_{-\infty}^{\infty} \!\!dx_b\, \avg{\,\ldots\,} .
\end{equation}
The respective moments (or other quantities of interest) will then carry the superscript ``$\,^*\,$".


Using the definition of local time, Eq.~\eqref{LTdef}, and the
Chapman-Kolmogorov formula
\begin{equation} \label{ChKolEq}
(x_b t_b | x_a t_a)
= \int_{-\infty}^{\infty} \!\!dx\, (x_b t_b | x\, t) (x\, t | x_a t_a) 
~,\quad t_a \leq t \leq t_b ,
\end{equation}
we can time-slice the path integral in \eqref{CorrFuns} to arrive at
\begin{align} \label{CorrFuns2}
\avg{L(x_1) \ldots L(x_n)}
&= \sum_{\sigma\in S_n} 
~~\int\displaylimits_{t_a < t_1 < \ldots < t_n < t_b} 
\!\!\!\!dt_1 \ldots dt_n
\nonumber\\
&\quad\times \prod_{k=0}^{n} 
(x_{\sigma(k+1)} t_{k+1} | x_{\sigma(k)} t_k ) ,
\end{align}
where $\sigma$ are permutations of indices $\{1,\ldots,n\}$, with $\sigma(0)=0$ and $\sigma(n+1)=n+1$, and we have denoted $x_0 = x_a$, $t_0 = t_a$, $x_{n+1} = x_b$ and $t_{n+1} = t_b$. 

Correlation functions provide an example of averages of functionals of the local time. In fact, the average of a generic functional $F[L(x)]$ can be neatly represented as
\begin{align} \label{FuncLTPI}
\avg{F[L(x)]} 
&= \int\displaylimits_{x(t_a)=x_a}^{x(t_b)=x_b} \hspace{-3mm} \mathcal{D}x \int \frac{\mathcal{D}p}{2\pi}\, e^{\mathcal{A}[p,x]} F[L(x)]
\nonumber\\
&= F\left[-\frac{\delta}{\delta U(x)}\right]
\int\displaylimits_{x(t_a)=x_a}^{x(t_b)=x_b} \hspace{-3mm} \mathcal{D}x \int \left. \frac{\mathcal{D}p}{2\pi}\, e^{\mathcal{A}_U[p,x]} \right|_{U=0} ,
\end{align}
where $\mathcal{A}_U[p,x]$ is the action corresponding to the Hamiltonian $H_U(p,x,t) = H(p,x,t) + U(x)$, where $U$ is an auxiliary external potential. The fact that the functional $F[L(x)]$ can be generated under the path integral sign by functional differentiation with respect to the ``source term" $U(x)$ (cf. the method of sources and generating functionals common in the quantum field theory \cite{Ramond}) is a consequence of the identity
\begin{equation} \label{PotLT}
\int_{t_a}^{t_b}  U(x(\tau)) \,d\tau
= \int_{-\infty}^{\infty} U(x) L(x) \,dx .
\end{equation}

As an example, let us fix points $x_1,\ldots,x_n$, and consider the functional $\prod_{j=1}^n \delta(L_j-L(x_j))$. Its normalized average yields the $n$-point distribution function of local time,
\begin{equation} \label{DistrLTn}
W(L_1,\ldots,L_n;x_1,\ldots,x_n)
= \frac{\bigavg{\prod_{j=1}^n\delta(L_j-L(x_j))}}{\avg{1}} .
\end{equation}


\section{Time-independent systems and the resolvent method}
\label{sec:TimeIndep}

Explicit calculations of correlation functions as well as of generic functional averages simplify significantly when we assume that the Hamiltonian $H$ is time-independent. In this case, the transition probabilities can be found by exponentiating the Hamiltonian operator, 
\begin{equation} \label{CondProbTimeIndep}
(x_b t_b | x_a t_a)
= \bra{x_b} e^{-(t_b-t_a)\hat{H}} \ket{x_a} .
\end{equation}
[We may now assume, without loss of generality, that $t_a=0$, and denote $t = t_b$.]

The Laplace transform with respect to the time variable expresses functional averages in the conjugate energy domain,
\begin{equation}
\avg{F[L(x)]}_E
= \int_{0}^{\infty} \!\!dt\, e^{-t E} \avg{F[L(x)]} .
\end{equation}
For correlation functions, this allows to carry out the convoluted integrals in Eq.~\eqref{CorrFuns2}, with a help of substitutions $t_k' = t_{k+1}-t_k$, and arrive at the formula
\begin{align} \label{npointE}
\avg{L(x_1) \ldots &L(x_n)}_E
= \sum_{\sigma\in S_n} \prod_{k=0}^n 
R(x_{\sigma(k)},x_{\sigma(k+1)},-E)
\end{align}
where
\begin{equation} \label{Resolvent}
R(x,x',-E) \equiv \bra{x'} (\hat{H}+E)^{-1} \ket{x}
\end{equation}
denotes the matrix elements of the resolvent operator ${\hat{R}(E)=(\hat{H}-E)^{-1}}$.

Knowing the resolvent, we can calculate, for example, the mean value of the local time at point $x$ by taking the Laplace inverse of the expression
\begin{equation} \label{1pointE}
\avg{L(x)}_E
= R(x_a,x) R(x,x_b) ,
\end{equation}
and by normalizing the result according to Eq.~\eqref{Moments}.
[We shall often abbreviate ${R(x,x') \equiv R(x,x',-E)}$.]
Likewise, the second moment can be obtained from
\begin{align} \label{2pointE}
\avg{L(x_1) L(x_2)}_E
&= R(x_a,x_1) R(x_1,x_2) R(x_2,x_b) \nonumber\\
&\quad+ R(x_a,x_2) R(x_2,x_1) R(x_1,x_b) .
\end{align}

For generic functionals of local time, the Laplace transform of Eq.~\eqref{FuncLTPI} yields
\begin{equation} \label{FuncLTRes}
\avg{F[L(x)]}_E
= \left. F\left[-\frac{\delta}{\delta U(x)}\right]
R_U(x_a,x_b,-E) \right|_{U=0} ,
\end{equation}
where $\hat{R}_U(E)=(\hat{H}_U - E)^{-1}$ is the resolvent corresponding to the extended Hamiltonian $H_U$. It satisfies the relation (sometimes termed the second resolvent identity)
\begin{equation} \label{ResIdElem}
R_U(x_a,x_b) 
= R(x_a,x_b) - \int_{-\infty}^{\infty} \!\!\!\!\!dx' R(x_a,x') U(x') R_U(x',x_b) ,
\end{equation}
which can be easily proved by operator partial fraction decomposition.

Finding the resolvent $R_U$ for a generic potential $U(x)$ is intractable. However, if we limit ourselves to functionals that only depend on the value of the local time at single point $x$, it is sufficient to consider the potential of the form
\begin{equation} \label{PotDelta}
U(x')=u\, \delta(x'-x) .
\end{equation}
For example, in order to calculate the one-point distribution function of local time at $x$ according to Eq.~\eqref{DistrLTn}, we need to find the Laplace inverse of the quantity
\begin{equation} \label{Distr1E}
\bigavg{\delta(L-L(x))}_E
= \int_{-\infty}^{\infty} \frac{ds}{2\pi} 
e^{-i s L} \avg{e^{i s L(x)}}_E 
\end{equation}
Under the assumption~\eqref{PotDelta}, variational derivatives with respect to $U$ at $x$ reduce to partial derivatives with respect to $u$, so that
\begin{equation}
\avg{e^{i s L(x)}}_E
= e^{-i s \partial_{u}} R_{U}(x_a,x_b) |_{u=0}
= R_U(x_a,x_b) |_{u=-i s}
\end{equation}

To find $R_U$, we use the identity~\eqref{ResIdElem}, which now takes the form
\begin{equation} \label{ResId1peak}
R_U(x_a,x_b) = R(x_a,x_b) - u R(x_a,x) R_U(x,x_b) 
\end{equation}
This can be solved by setting ${x_a=x}$, expressing $R_U(x,x_b)$ in terms of the ``free" resolvent $R$, and substituting back to Eq.~\eqref{ResId1peak}, with the result
\begin{equation}
R_U(x_a,x_b) 
= R(x_a,x_b) - u \frac{R(x_a,x) R(x,x_b)}{1+u R(x,x)} .
\end{equation}
Putting $u=-is$, and calculating the Fourier integral \eqref{Distr1E} with a help of Formulas~\eqref{FourTr1} and \eqref{FourTr2}, valid for ${\rm Re}\,R(x,x)>0$, we obtain the result
\begin{align} \label{1pointDistE}
\bigavg{\delta(L-&L(x))}_E
= \theta(L) \frac{R(x_a,x) R(x,x_b)}{R(x,x)^2}  
e^{-\frac{L}{R(x,x)} } \nonumber\\
&+ \delta(L) \left[ R(x_a,x_b) - \frac{R(x_a,x) R(x,x_b)}{R(x,x)} \right] .
\end{align}
Here, the step function $\theta$ confirms the non-negativity of local time, and the $\delta$-peak signals concentration of some part of the probability measure at the origin of the local-time axis. The latter phenomenon can be intuitively understood as follows. If the point $x$ does not lie between the two endpoints $x_a$ and $x_b$, then there is a substantial fraction of paths that never cross $x$, and therefore have zero local time at $x$. Note that for $x=x_a$ or $x_b$, the $\delta$-term vanishes.



To look for multipoint distributions, one needs to consider the potential $U$ with several $\delta$-peaks. A generalization of the previous method  is straightforward, however, the complexity of calculations rapidly increases (see Appendix~\ref{sec:MultiDist}).

Before moving to the next section let us shortly comment on the possibility of investigating local times in higher-dimensional position spaces. There, the local-time profile corresponding to a generic sample path ${\bf x}(\tau)$ is a highly singular function of ${\bf x}$. Nevertheless, our manipulations in this section still make sense, and the results for averaged quantities, in particular, Eqs.~\eqref{npointE} and \eqref{1pointDistE}, readily generalize to the multidimensional case.

\section{Local time of L\'{e}vy random walks}
\label{sec:Levy}

In this section we will assume that the Hamiltonian is, in addition to its time-independence, also translationally invariant, i.e., $H$ is a function of the momentum $p$ only. 

Using the resolution of identity in the momentum basis, ${\hat{1} = \int \!dp \proj{p}}$, where ${\braket{x}{p} = e^{i p x}/\sqrt{2\pi}}$, we can express the transition elements~\eqref{CondProbTimeIndep}, which  depend only on the difference ${x=x_b-x_a}$ of the two endpoints, as
\begin{equation}
P(x,t)
= (x_b t | x_a 0)
= \int_{-\infty}^{\infty} \frac{dp}{2\pi}
e^{-t H(p)} e^{i p x} .
\end{equation}
$P(x,t)$ is the distribution after time $t$ of a particle that starts at the origin, and undergoes a random walk described by the Hamiltonian $H(p)$. To ensure proper normalization, i.e., that $\avg{1}^*=1$, we will assume $H(0)=0$. 
[Note that $e^{-t H(-p)}$ is the characteristic function of the probability distribution $P(x)$.]

In the momentum representation, the resolvent corresponding to $H(p)$ reads
\begin{equation} \label{ResHp}
R(x,x',-E) 
= \int_{-\infty}^{\infty} \frac{dp}{2\pi}
\frac{e^{i p (x'-x)}}{H(p) + E} .
\end{equation}
Occasionally, we will use the fact that $R$ is a function of difference of the two positions to write ${R(x'-x)=R(x,x')}$.
Let us remark that integration over an endpoint yields
\begin{equation} \label{ResIntegrated}
\int_{-\infty}^{\infty} \!\!dx'\, R(x,x',-E)
= \frac{1}{E} ,
\end{equation}
regardless of the Hamiltonian $H(p)$.

\begin{figure}
\includegraphics[scale=1]{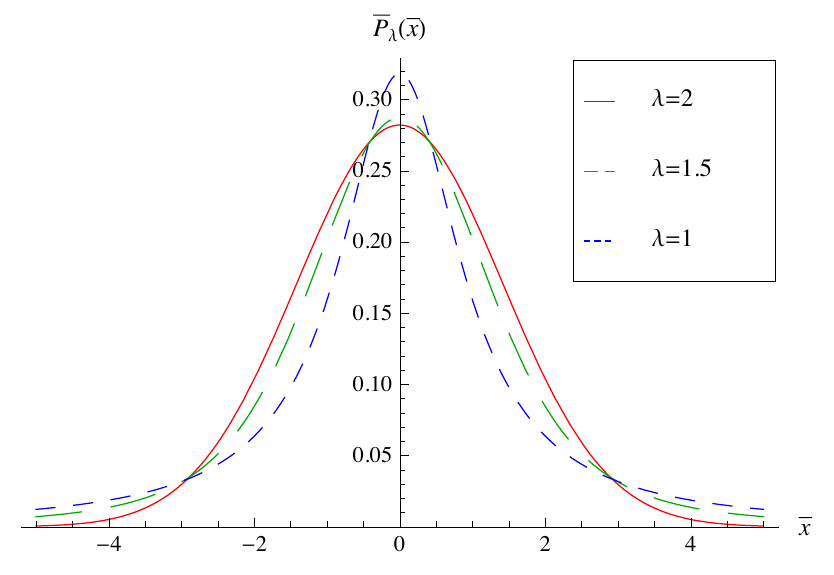} 
\caption{The L\'{e}vy stable distribution $P_\lambda(x,t)$ for $\lambda=2$ (Gaussian), $\lambda=1.5$, and $\lambda=1$ (Cauchy). Dimensionless quantity $\oline{P}_\lambda(\oline{x})=(D_\lambda t)^{1/\lambda} P_\lambda(x,t)$ is plotted, where $\oline{x}=x/(D_\lambda t)^{1/\lambda}$.}
\label{fig:PlotDistr}
\end{figure}

From now on, we shall focus on the L\'{e}vy Hamiltonian~\eqref{LevyHam}, for which the distribution of the walk is given by the symmetric L\'{e}vy stable distribution (see Fig.~\ref{fig:PlotDistr})
\begin{equation} \label{StableDist}
P_\lambda(x,t)
= \int_{-\infty}^{\infty} \frac{dp}{2\pi}
e^{-t D_\lambda (p^2)^{\lambda/2}} e^{i p x} .
\end{equation}
This reduces to the Gaussian distribution for $\lambda=2$,
\begin{equation} \label{GaussDist}
P_2(x,t)
= \frac{e^{-\frac{x^2}{4 D_2 t}}}{\sqrt{4\pi D_2 t}} ,
\end{equation}
and to the Cauchy(-Lorentz) distribution for $\lambda=1$,
\begin{equation} \label{CauchyDist}
P_1(x,t)
= \frac{1}{\pi} \frac{D_1 t}{(D_1 t)^2 + x^2} .
\end{equation}

For generic $\lambda\in(1,2)$, the integral over $p$ is not tractable. We can only find the value of $P_\lambda$ at the origin (the recurrence probability)
\begin{equation} \label{RecurProb}
P_\lambda(0,t)
= \frac{\Gamma(\frac{1}{\lambda})}
{\lambda\pi (D_\lambda t)^{1/\lambda}} ,
\end{equation}
and the asymptotic behaviour 
\begin{equation} \label{LevyTail}
P_\lambda(x,t)
\overset{|x|\rightarrow\infty}{\sim}
\frac{D_\lambda t}{2\pi |x|^{1+\lambda}}
\Gamma(1+\lambda) \sin\frac{\pi\lambda}{2} .
\end{equation}
We observe that for $\lambda<2$, the L\'{e}vy distribution develops heavy tails, i.e., follows a power law for large $|x|$. In the Gaussian case $\lambda=2$, the numerical factor on the right-hand side of Eq.~\eqref{LevyTail} vanishes, in agreement with the exponential decay of the normal distribution.

\begin{figure}
\includegraphics[scale=1]{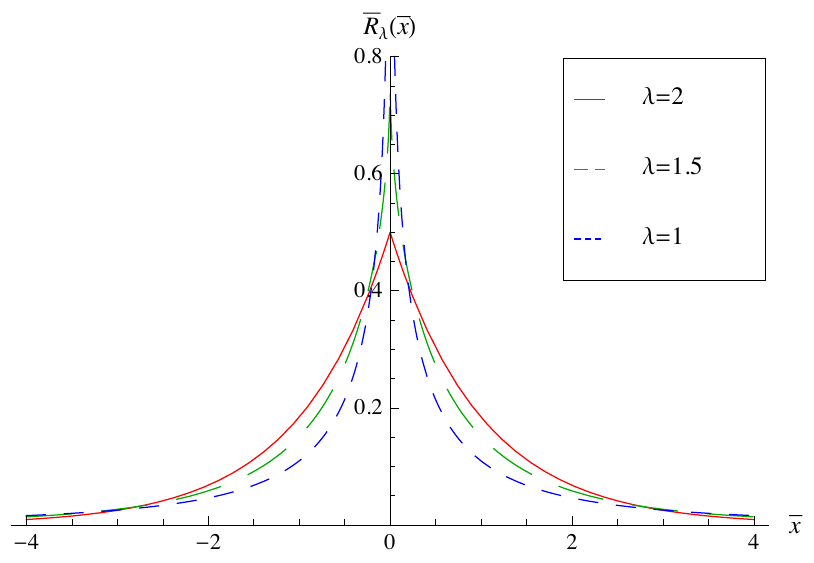} 
\caption{The resolvent $R_\lambda(x,-E)$ for $\lambda=1,1.5,2$. Dimensionless quantity $\oline{R}_\lambda(\oline{x})=(D_\lambda/E)^{1/\lambda} E R_\lambda(x,-E)$ is plotted, where $\oline{x}=x/(D_\lambda/E)^{1/\lambda}$.}
\label{fig:PlotRes}
\end{figure}

The resolvent of the Hamiltonian $H_\lambda$,
\begin{align} \label{ResLevy}
R_\lambda(x,-E)
&=  \int_{-\infty}^{\infty} \frac{dp}{2\pi}
\frac{e^{i p x}}{D_\lambda (p^2)^{\lambda/2} + E} ,
\end{align} 
is known in mathematics as the Linnik (or geometric stable) distribution \cite{Linnik1953}, \cite[Ch.~4.3]{Kotz}. It is plotted in Fig.~\ref{fig:PlotRes}. 
For $\lambda=2$ we obtain the ``Gaussian" resolvent
\begin{equation} \label{ResGaussian}
R_2(x,-E)
= \frac{e^{-\sqrt{E/D_2} |x| }}{\sqrt{4 D_2 E}} ,
\end{equation}
and for $\lambda=1$ the ``Cauchy" resolvent
\begin{equation}
R_1(x,-E)
= \frac{-1}{\pi D_1} \left[ 
\sin\frac{E|x|}{D_1} \, {\rm si}\frac{E |x|}{D_1}
+ \cos\frac{E|x|}{D_1} \, {\rm ci}\frac{E|x|}{D_1} \right] ,
\end{equation}
where the sine and cosine integrals are defined in Appendix~\ref{sec:Formulas}, Eq.~\eqref{SiCi}.
For generic $\lambda$, we may find the diagonal elements of the resolvent
\begin{equation} \label{ResDiag}
R_\lambda(0,-E)
= \frac{E^{\frac{1}{\lambda}-1}}{ \lambda D_\lambda^{1/\lambda} \sin\frac{\pi}{\lambda}} .
\end{equation}
The tail behaviour follows easily from Eq.~\eqref{LevyTail} and the relation 
\begin{equation}
R_\lambda(x,-E)
= \int_{0}^{\infty} \!\!dt\, e^{-t E} P_\lambda(x,t) 
\end{equation}
between the transition probabilities and the resolvent.

Finally, let us also mention an alternative representation of $R_\lambda$ obtained by rotating the momentum axis to imaginary values \cite{Kotz1995},
\begin{equation}
R_\lambda(x,-E)
= -\frac{1}{\pi} {\rm Im} \int_{0}^{\infty} \!\!dv\,
\frac{e^{-v|x|}}{E + D_\lambda v^\lambda e^{i \pi \lambda / 2}} ,
\end{equation}
which expresses a generic Linnik distribution as a superposition of Gaussian resolvents~\eqref{ResGaussian}.

In Fig.~\ref{fig:SamplePaths} we plot sample paths of L\'{e}vy random walks for various values of the parameter $\lambda$. We observe that for decreasing $\lambda$ the paths tend to develop characteristic long jumps. These are then reflected in the corresponding local-time profiles, which consist of more and more isolated peaks.

\subsection{One-point distribution: generic $\lambda$}

For simplicity, we will consider one-point distributions of the local time only at the initial point $x=x_a$, in which case the second term in Eq.~\eqref{1pointDistE} vanishes. We will also assume that $L \geq 0$ to remove the step function $\theta(L)$. 

Two cases, which allow further simplifications, will be considered --- the set of paths with fixed endpoint ${x_b=x_a}$, and the set of paths with arbitrary endpoint (see Eq.~\eqref{EndpointInt}). 

\begin{figure}
\includegraphics[scale=1]{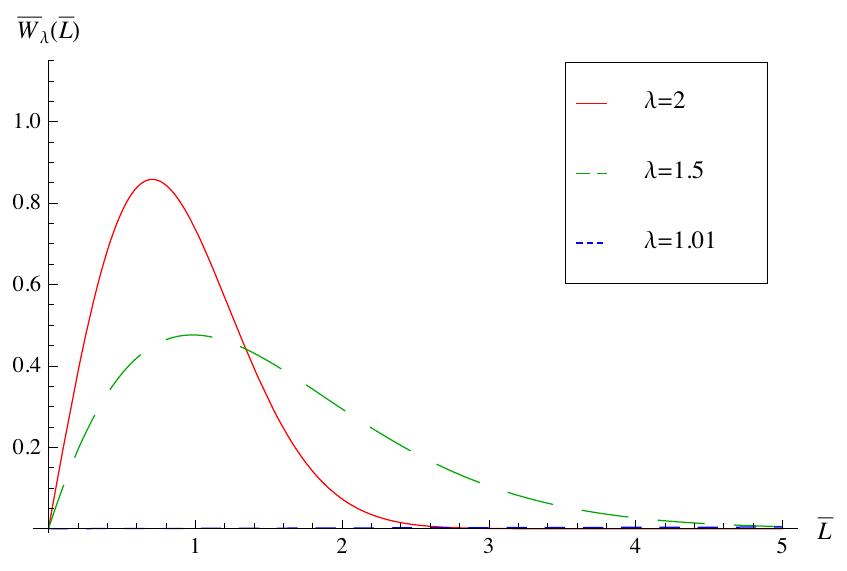}  
\caption{One-point distribution of local time at point $x=x_a$ for fixed endpoint $x_b=x_a$. Dimensionless quantity $\overline{W}_\lambda(\overline{L}) = t \,W_\lambda(L;x_a)/(D_\lambda t)^{1/\lambda}$ is plotted, where $\overline{L}=(D_\lambda t)^{1/\lambda}L/t$.}
\label{fig:LTDistrXa}
\end{figure}

In the first case, Eq.~\eqref{1pointDistE} requires only a knowledge of the diagonal matrix elements of the resolvent $R_\lambda$, registered in Eq.~\eqref{ResDiag}:
\begin{equation} \label{Distr1XaE}
\bigavg{\delta(L-L(x_a))}_E^{x_b=x_a}
= e^{-\frac{L}{R_\lambda(0)} }
= \exp\left[- \sigma_\lambda L E^{1-\frac{1}{\lambda}} \right] ,
\end{equation}
where we have denoted 
\begin{equation}
\sigma_\lambda \equiv \lambda \, D_\lambda^{1/\lambda} \sin\frac{\pi}{\lambda} .
\end{equation}
The inverse Laplace transform in $E$ can be represented by the Bromwich integral, where we place the branch cut along the negative $E$ axis, and deform the integration contour so as to enclose it. Normalizing the result by the factor $\avg{1}=P_\lambda(0,t)$, given by Eq.~\eqref{RecurProb}, we obtain an integral representation for the distribution of local time at point $x_a$, for $x_b=x_a$,
\begin{align}
W_\lambda(L;x_a)
= \frac{\lambda (D_\lambda t)^{1/\lambda}}{\Gamma(\frac{1}{\lambda})}
&\int_0^\infty \!\!\! dE \, e^{-E t} 
\exp\left[L E^{1-\frac{1}{\lambda}} \sigma_\lambda \cos\frac{\pi}{\lambda} \right]
\nonumber\\
&\quad\times \sin\left[ L E^{1-\frac{1}{\lambda}} \sigma_\lambda \sin\frac{\pi}{\lambda} \right] ,
\end{align}
plotted in Fig.~\ref{fig:LTDistrXa}.
For $\lambda=2$, the integral can be calculated explicitly:
\begin{equation}
W_2(L;x_a) 
= \frac{2 D_2 L}{t} e^{-D_2 L^2/t} .
\end{equation}

\begin{figure}
\includegraphics[scale=1]{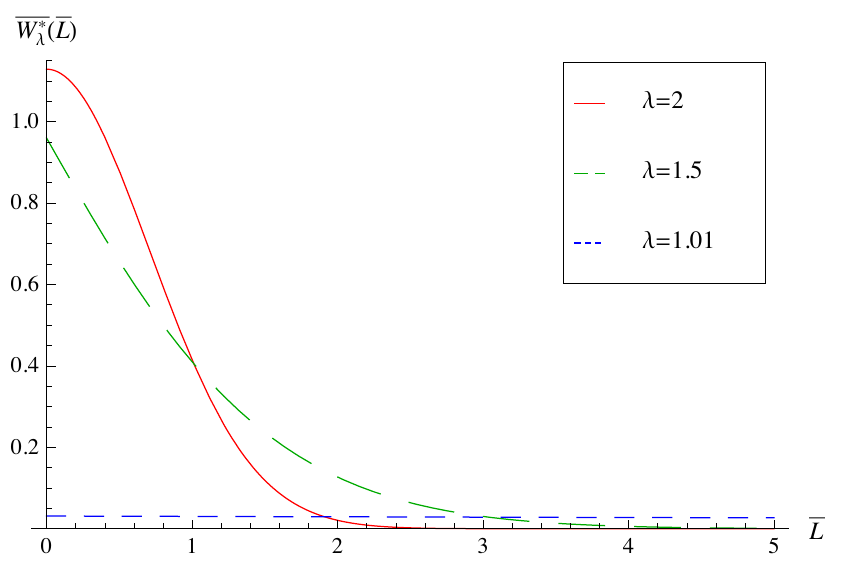} 
\caption{One-point distribution of local time at point $x=x_a$ for paths with arbitrary $x_b$. Dimensionless quantity $\overline{W_\lambda^*}(\overline{L}) = t \,W_\lambda^*(L;x_a)/(D_\lambda t)^{1/\lambda}$ is plotted, where $\overline{L}=(D_\lambda t)^{1/\lambda}L/t$.}
\label{fig:LTDistrInt}
\end{figure}

In the second case of interest we integrate Eq.~\eqref{1pointDistE} over the endpoint $x_b$, and find, due to the property~\eqref{ResIntegrated},
\begin{equation} \label{Distr1IntE}
\bigavg{\delta(L-L(x_a))}_E^{*}
= \frac{ e^{-\frac{L}{R_\lambda(0)} } }{E R_\lambda(0)}
= \frac{\sigma_\lambda}{ E^{1/\lambda}}
\exp\left[- \sigma_\lambda L E^{1-\frac{1}{\lambda}} \right] .
\end{equation}
Inverting the Laplace transform as in the previous case yields the one-point distribution of local time at point $x_a$ for integrated $x_b$
\begin{align}
W_\lambda^*(L;x_a) 
= \frac{\sigma_\lambda}{\pi}
&\int_0^\infty \!\! dE \, \frac{e^{-E t}}{E^{1/\lambda}} 
\exp\left[L E^{1-\frac{1}{\lambda}} \sigma_\lambda \cos\frac{\pi}{\lambda} \right]
\nonumber\\
&\quad\times \sin\left[ L E^{1-\frac{1}{\lambda}} \sigma_\lambda \sin\frac{\pi}{\lambda} + \frac{\pi}{\lambda} \right] ,
\end{align}
depicted in Fig.~\ref{fig:LTDistrInt}.
For the Gaussian case $\lambda=2$, this reduces to
\begin{equation}
W_2^*(L;x_a)
= \sqrt{\frac{4 D_2}{\pi t}} \,e^{-D_2 L^2/t} .
\end{equation}

Note that the one-point distributions $W_\lambda$ and $W_\lambda^*$ depend quite significantly on the L\'{e}vy index $\lambda$. This suggests a local-time-based method for identifying $\lambda$ by making statistics of the number of times the stochastic trajectories hit the origin.

\subsection{One-point distribution: Cauchy case $\lambda=1$}

When $\lambda$ approaches $1$, we observe an interesting phenomenon of flattening of the local-time one-point distributions (see Figs.~\ref{fig:LTDistrXa} and \ref{fig:LTDistrInt}). This is due to the fact that in the Cauchy limit $\lambda=1$, the resolvent becomes singular on the diagonal, i.e.,
\begin{equation}
\lim_{\lambda \searrow 1}
R_\lambda(0,-E)
= +\infty .
\end{equation}
In effect, the exponent on the right-hand side of Eqs.~\eqref{Distr1XaE} and \eqref{Distr1IntE} vanishes, and so the one-point distribution functions ceases to depend on $L$. That is, for $\lambda$ close to $1$, a sample path is equally likely to spend small and large amounts of time in the vicinity of the point $x_a$.

\subsection{One-point distribution: Gaussian case $\lambda=2$}

For $\lambda=2$, the resolvent assumes a particularly simple form~\eqref{ResGaussian}, which allows us to calculate one-point local-time distributions at arbitrary point, and for the final point $x_b$ not necessarily equal to the initial point $x_a$. Substituting \eqref{ResGaussian} into Eq.~\eqref{1pointDistE} we obtain
\begin{widetext}
\begin{align} \label{1pointEGauss}
\bigavg{\delta(L-L(x))}_E 
&= \theta(L) \, \exp\left[-\sqrt{\frac{E}{D_2}} \big(|x_a-x|+|x-x_b|+2 D_2 L \big) \right]
\nonumber\\
&\quad + \frac{\delta(L)}{\sqrt{4 D_2 E}}
\left\{ \exp\left[ -\sqrt{\frac{E}{D_2}} \,|x_b-x_a| \right] 
- \exp\left[-\sqrt{\frac{E}{D_2}}\big(|x_a-x|+|x-x_b| \big) \right] \right\}
\end{align}

Performing inverse Laplace transform with a help of \cite[Ch.~17.13]{Gradshteyn}, and normalizing by $\avg{1} = P_2(x_b-x_a,t)$, we find the distribution of Brownian local time, i.e., the local time corresponding to the Brownian motion, at point $x$ for fixed final point $x_b$
\begin{align} \label{1pointGauss}
W_2(L;x)
&= \theta(L)
\frac{|x_a-x|+|x-x_b|+2 D_2 L}{t} 
\exp\left[ -\frac{(|x_a-x|+|x-x_b|+2 D_2 L)^2 - (x_b-x_a)^2}{4 D_2 t} \right] \nonumber\\
&\quad + \delta(L)
\left\{ 1 - 
 \exp\left[ -\frac{(|x_a-x|+|x-x_b|)^2 - (x_b-x_a)^2}{4 D_2 t} \right]
\right\} ,
\end{align}
\end{widetext}
recovering the results of \cite[Ch.~I.2]{Borodin1989}. Note that the $\delta$-term vanishes whenever $x$ lies between $x_a$ and $x_b$.

For paths with unspecified endpoint we integrate Eq.~\eqref{1pointEGauss} over $x_b$, and invert the Laplace transform [recall that the normalization factor $\avg{1}^*$ is equal to $1$] to find the distribution of Brownian local time at point $x$ for ``free" final point
\begin{align}
W_2^*(L;x)
&= \theta(L) \sqrt{\frac{4 D_2}{\pi t}} 
\exp\left[ -\frac{(|x_a-x|+2 D_2 L)^2}{4 D_2 t} \right]
\nonumber\\
&\quad + \delta(L) \, {\rm erf}\left[ \frac{|x_a-x|}{\sqrt{4 D_2 t}} \right] ,
\end{align}
where the error function ${\rm erf}(x)$ is defined in the standard way by Eq.~\eqref{ErrorFun}.
This result can be found already in \cite{Levy1965}.

\subsection{First moment: generic $\lambda$}

Now, let us investigate the first moment (or the mean) of the local time distribution at a given point $x$ as a function of $x$. The following manipulations hold for an arbitrary Hamiltonian $H(p)$.

There are two ways how to proceed.
First, we can start from the one-point ``correlation" function in the $E$-domain, Eq.~\eqref{1pointE}, which for the resolvent~\eqref{ResHp} reads
\begin{equation}
\avg{L(x)}_E
= \int_{-\infty}^{\infty} 
\frac{dp}{2\pi} \frac{dp'}{2\pi}
\frac{e^{i p (x-x_a)}}{H(p) + E} 
\frac{e^{i p' (x_b-x)}}{H(p') + E} .
\end{equation}
In the time domain, we then find the general formula 
\begin{equation}
\mu_{}(x)
= \int_{-\infty}^{\infty} 
\frac{dp}{2\pi} \frac{dp'}{2\pi}
\frac{e^{-t H(p)} - e^{-t H(p')}}{H(p') - H(p)}
\frac{e^{i p (x-x_a)} e^{i p' (x_b-x)}}{P(x_b-x_a,t)}
\end{equation}
for the mean of the local time at point $x$ for fixed endpoint $x_b$.
Likewise, by integration over $x_b$ one obtains the mean of the local time
\begin{equation}
\mu_{}^*(x) 
= \int_{-\infty}^{\infty} 
\frac{dp}{2\pi}
\,\frac{1 - e^{-t H(p)}}{H(p)} \,e^{i p (x-x_a)}
\end{equation}
for unspecified endpoint $x_b$.

Alternatively, one can use the formula~\eqref{CorrFuns2} to find
\begin{equation} \label{MeanP}
\mu(x)
= \int_0^t \!\!dt_1\, 
\frac{P(x_b-x,t-t_1) P(x-x_a,t_1)}{P(x_b-x_a,t)} ,
\end{equation}
and
\begin{equation} \label{MeanPInt}
\mu^*(x)
= \int_0^t \!\!dt_1\, P(x-x_a,t_1) ,
\end{equation}
where we have used the fact that $P(x)$ is normalized to $1$.

\begin{figure}
\includegraphics[scale=1]{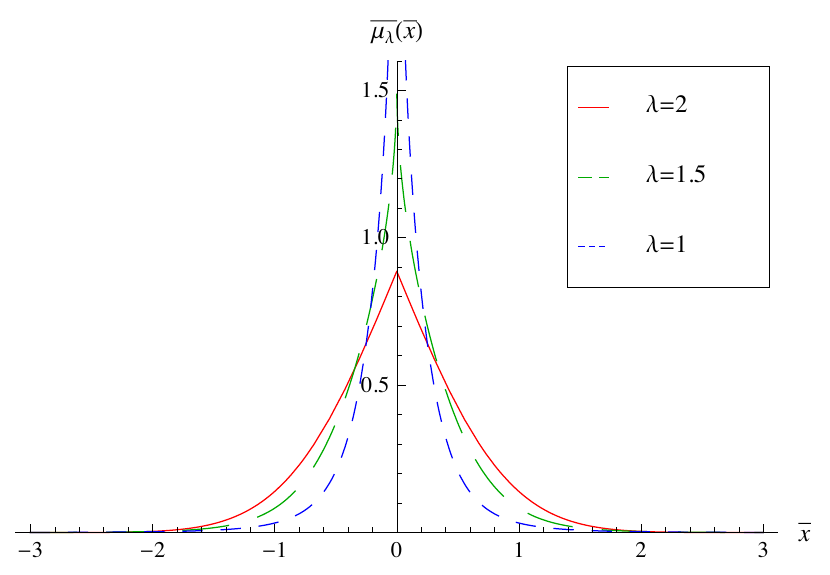}  
\caption{The mean of the local time at point $x$ for the L\'{e}vy Hamiltonian $H_\lambda$ and fixed endpoint ${x_b=x_a}$. Dimensionless quantity $\overline{\mu_\lambda}(\overline{x})=(D_\lambda t)^{1/\lambda}\mu_\lambda(x)/t$ is plotted, where ${\overline{x}=(x-x_a)/(D_\lambda t)^{1/\lambda}}$.}
\label{fig:LTMom1Xa}
\end{figure}

\begin{figure}
\includegraphics[scale=1]{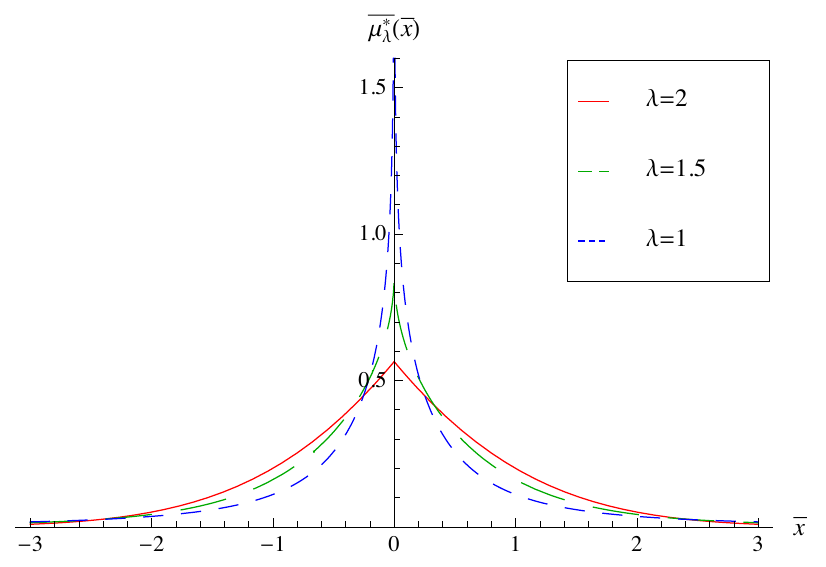} 
\caption{The mean of the local time at point $x$ for the L\'{e}vy Hamiltonian $H_\lambda$ and ``free" endpoint $x_b$. Dimensionless quantity $\overline{\mu_\lambda^*}(\overline{x})=(D_\lambda t)^{1/\lambda}\mu_\lambda^*(x)/t$ is plotted, where ${\overline{x}=(x-x_a)/(D_\lambda t)^{1/\lambda}}$.}
\label{fig:LTMom1Int}
\end{figure}

For the case of L\'{e}vy Hamiltonian $H_\lambda$, the first moments $\mu(x)$ and $\mu^*(x)$ are plotted in Figs.~\ref{fig:LTMom1Xa} and \ref{fig:LTMom1Int} for various values of $\lambda$.


Formulas~\eqref{MeanP} and \eqref{MeanPInt} are particularly useful when one is interested in the tail behaviour of the first moment for large $x$, which can be easily investigated with a help of Formula~\eqref{LevyTail}, or when the transition probabilities are know explicitly, as will be the case in the following section.

\subsection{First moment: Cauchy case $\lambda=1$}

For the case of Cauchy random walk, characterized by the value $\lambda=1$, one can use the explicit result~\eqref{CauchyDist} in Eq.~\eqref{MeanP} to arrive at 
\begin{align}
&\mu_1(x)
= 
\frac{-1}{2 \pi D_1^2  t \left[ (D_1 t)^2 + (x_a+x_b-2 x)^2 \right]}
\nonumber\\
&\!\!\times\!
\left\{
2 (x \!-\! x_a)\! \left[ (D_1 t)^2 \!+\! (x \!-\! x_a)^2 \!-\! (x_b \!-\! x)^2 \right] 
\arctan\frac{D_1 t}{x - x_a} \right.
\nonumber\\
&+ 2 (x_b \!-\! x)\! \left[ (D_1 t)^2 \!+\! (x_b \!-\! x)^2 \!-\! (x \!-\! x_a)^2 \right] 
\arctan\frac{D_1 t}{x_b - x}
\nonumber\\
&+ D_1 t \left[ (D_1 t)^2 + (x - x_a)^2 + (x_b - x)^2 \right]
\nonumber\\
&\quad\times
\left.\ln\frac{(x - x_a)^2 (x_b - x)^2}
{\left[ (D_1 t)^2 + (x - x_a)^2 \right] 
\left[ (D_1 t)^2 + (x_b - x)^2) \right]} \right\} ,
\end{align}
and in Eq.~\eqref{MeanPInt} to find
\begin{equation}
\mu_{1}^*(x) 
= \frac{1}{2 \pi D_1}
\ln\left[ 1 + \left(\frac{D_1 t}{x-x_a}\right)^2 \right] .
\end{equation}
These functions both diverge when $x$ approaches $x_a$.

\subsection{First moment: Gaussian case $\lambda=2$}

For the Gaussian random walk (the Brownian motion), it is convenient to use Formula~\eqref{1pointE} with the Gaussian resolvent~\eqref{ResGaussian}. The first moments of the local time are then found in a straightforward manner with a help of \cite[Ch.~17.13]{Gradshteyn}:
\begin{equation}
\mu_{2}(x)
= \sqrt{\frac{\pi t}{4 D_2}} 
\,e^{\frac{(x_b-x_a)^2}{4 D_2 t} }
\,{\rm erfc}\left[ \frac{|x_a-x|+|x-x_b|}{\sqrt{4 D_2 t}} \right]
\end{equation}
for fixed $x_b$, and
\begin{equation}
\mu_2^*(x) 
= \sqrt{\frac{t}{\pi D_2}} \,e^{- \frac{(x_a-x)^2}{4 D_2 t} }
- \frac{|x_a-x|}{2 D_2} \,{\rm erfc}\left[ \frac{|x_a-x|}{\sqrt{4 D_2 t}} \right]
\end{equation}
for integrated $x_b$.

Second moments of Brownian local time are calculated in Appendix~\ref{sec:HigherMom}.

\section{Conclusion and outlook}

In this article, we have studied local times of stochastic processes corresponding to the diffusion equation~\eqref{HeatEq}. We believe that our path-integral-based approach, combined with operator methods and notation of standard quantum mechanics, is more accessible to physicists than usual mathematical random walk treatments \cite{MarcusRosen,Borodin1989}. Moreover, the technique of variational differentiation with respect to an auxiliary ``source" (in our case the auxiliary potential $U$), extensively used in the quantum field theory, provides an elegant path-integral representation~\eqref{FuncLTPI} of generic functionals of the local time, including moments, correlations, and local time distributions.

For time-independent systems, i.e., those whose evolution is generated by a time-independent Hamiltonian $H(p,x)$, we have revealed an intimate connection between the local time and the resolvent of the Hamiltonian operator $\hat{H}$. 
This connection was substantiated in Formulas~\eqref{npointE} for correlation functions (from which the moments of the local time easily follow) and \eqref{FuncLTRes} for generic functionals of the local time. The latter result allowed us to calculate distribution functions of the local time (see Eqs.~\eqref{1pointDistE} and Appendix~\ref{sec:MultiDist}).

Special attention has been devoted to L\'{e}vy random walks (or, L\'{e}vy flights), generated by the Hamiltonian~\eqref{LevyHam}. In Sec.~\ref{sec:Levy}, we have calculated one-point distribution functions and the first moments of the local time for the L\'{e}vy index $\lambda\in[1,2]$, thus extending existing results for the Brownian local time~\cite{Borodin1989}.

Higher moments and multipoint distribution functions of the local time deserve further investigation, although the matter is anticipated to be technically rather involved. The main tools and strategies have been outlined in this article, and illustrated with simple one-point cases.


\appendix

\section{Multipoint distributions}
\label{sec:MultiDist}

In order to calculate $n$-point distribution functions of local time, defined by Eq.~\eqref{DistrLTn}, we need to take the inverse Laplace transform of the quantity
\begin{align} \label{MultiPointE}
\bigavg{\prod_{j=1}^n \delta(L_j-L(x_j))}_E 
&= \prod_{k=1}^n \left[ \int_{-\infty}^{\infty}\frac{ds_k}{2\pi} e^{-i s_k L_k} \right]
\nonumber\\
&\quad\times
\bigavg{\prod_{j=1}^n e^{i s_j L(x_j)}}_E \,.
\end{align}
In analogy with Eq.~\eqref{PotDelta}, we assume the source potential of the form
\begin{equation} \label{PotMultiDelta}
U(x) = \sum_{j=1}^n u_j \delta(x-x_j) ,
\end{equation}
and find
\begin{equation}
\bigavg{\prod_{j=1}^n e^{i s_j L(x_j)}}_E
= R_U(x_a,x_b) |_{u_1=-i s_1,\ldots,u_n=-i s_n} .
\end{equation}

In this case, the resolvent identity~\eqref{ResIdElem} reduces to
\begin{equation} \label{ResIdnpeaks}
R_U(x_a,x_b) = R(x_a,x_b) - \sum_{j=1}^n u_j R(x_a,x_j) R_U(x_j,x_b) .
\end{equation}
Setting $x_a=x_k$ for $k=1,\ldots,n$, we obtain $n$ equations
\begin{equation}
\sum_{j=1}^n \big( \delta_{jk} + u_j R(x_k,x_j) \big) R_U(x_j,x_b)
= R(x_k,x_b) ,
\end{equation}
from which 
\begin{equation}
R_U(x_j,x_b) = \sum_{k=1}^n M_{j k}^{-1} R(x_k,x_b) ,
\end{equation}
where $M_{j k}^{-1}$ are elements of the inverse of the matrix
\begin{equation}
M_{jk} = \delta_{jk} + u_k R(x_j,x_k) . 
\end{equation}
Substituting back to Eq.~\eqref{ResIdnpeaks}, we obtain
\begin{equation}
R_U(x_a,x_b) 
= R(x_a,x_b) - \sum_{j,k=1}^n u_j R(x_a,x_j) M_{j k}^{-1} R(x_k,x_b) .
\end{equation}

This form of $R_U$ allows one, in principle, to perform the Fourier transforms in Eq.~\eqref{MultiPointE}. 

%

\section{Second moments}
\label{sec:HigherMom}

We shall only focus on the local time of Brownian motion, i.e., of a random walk generated by the Hamiltonian~\eqref{LevyHam} with $\lambda=2$.

Eq.~\eqref{2pointE} for the two-point correlation function with Gaussian resolvent~\eqref{ResGaussian} reads
\begin{equation} \label{2pointEGauss}
\avg{L(x_1) L(x_2)}_E
= \frac{e^{-\sqrt{E/D_2} \,\xi_{12}} + e^{-\sqrt{E/D_2} \,\xi_{21}}}{(4 D_2 E)^{3/2}} ,
\end{equation}
where we have denoted
\begin{align}
\xi_{12} 
\equiv |x_a-x_1| + |x_1-x_2| + |x_2-x_b| , \nonumber\\
\xi_{21} 
\equiv |x_a-x_2| + |x_2-x_1| + |x_1-x_b| .
\end{align}
From this we obtain the second moment of local time at points $x_1$ and $x_2$ (for fixed endpoint $x_b$)
\begin{align}
\mu_2(x_1,x_2)
&= \frac{t}{2 D_2} \left\{
\exp\left[- \frac{\xi_{12}^2 - (x_b-x_a)^2}{4 D_2 t} \right] \right.
\nonumber\\
&\quad - \left. \frac{\sqrt{\pi} \,\xi_{12}}{\sqrt{4 D_2 t}} 
\, {\rm erfc}\left[ \frac{\xi_{12}}{\sqrt{4 D_2 t}} \right]
\, \exp\left[ \frac{(x_b-x_a)^2}{4 D_2 t} \right] \right\}
\nonumber\\
&+ (\xi_{12} \leftrightarrow \xi_{21})
\end{align}

Integrating Eq.~\eqref{2pointEGauss} over $x_b$, and making use of Eq.~\eqref{ResIntegrated}, we find
\begin{equation}
\avg{L(x_1) L(x_2)}_E^*
= \frac{e^{-\sqrt{E/D_2} \,\xi_{12}^*} + e^{-\sqrt{E/D_2} \,\xi_{21}^*}}{4 D_2 E^2} ,
\end{equation}
where
\begin{align}
\xi_{12}^*
\equiv |x_a-x_1| + |x_1-x_2| , \nonumber\\
\xi_{21}^* 
\equiv |x_a-x_2| + |x_2-x_1| ,
\end{align}
and hence the second moment of the local time (for free endpoint $x_b$)
\begin{align}
\mu_2^*(x_1,x_2)
&= \frac{1}{4 D_2}
\left\{
-\frac{\sqrt{t} \,\xi_{12}^*}{\sqrt{\pi D_2}}
\exp\left[ -\frac{(\xi_{12}^*)^2}{4 D_2 t} \right] \right.
\nonumber\\
&\quad + \left. \left[ \frac{(\xi_{12}^*)^2}{2 D_2} + t \right]
{\rm erfc}\left[ \frac{\xi_{12}^*}{\sqrt{4 D_2 t}} \right]
\right\}
\nonumber\\
&+ (\xi_{12}^* \leftrightarrow \xi_{21}^*) .
\end{align}

\section{List of definitions and formulas}
\label{sec:Formulas}



We define the sine and cosine integrals in accordance with \cite[3.721]{Gradshteyn},
\begin{align} \label{SiCi}
{\rm si}(x) = -\int_{1}^{\infty} \frac{\sin(p x)}{p} \,dp \nonumber\\
{\rm ci}(x) = -\int_{1}^{\infty} \frac{\cos(p x)}{p} \,dp .
\end{align}

The error function and its complementary are also defined in a standard way,
\begin{align} \label{ErrorFun}
{\rm erf}(x)
&= \frac{2}{\sqrt{\pi}} \int_0^x e^{-u^2} du \nonumber\\
{\rm erfc}(x) 
&= 1 - {\rm erf}(x) .
\end{align}

In addition, let us quote some formulas that find applications in the main text. By methods of complex contour integration, one can easily verify that (for ${\rm Re}\,R>0$)
\begin{equation} \label{FourTr1}
\int_{-\infty}^{\infty} \frac{ds}{2\pi} 
\frac{e^{-i s L}}{1-i s R}
= \frac{e^{-\frac{L}{R}} \theta(L)}{R} .
\end{equation}
By differentiation with respect to $L$ one further obtains
\begin{equation} \label{FourTr2}
\int_{-\infty}^{\infty} \frac{ds}{2\pi} 
\frac{i s \,e^{-i s L}}{1-i s R}
= \frac{e^{-\frac{L}{R}} \theta(L)}{R^2} - \frac{\delta(L)}{R} .
\end{equation}

\end{document}